\documentstyle[12pt,epsfig,rotating]{article}
\def\pge{\pagestyle{empty}} \def\pgn{\pagestyle{plain}}
  
\def\spg{\setcounter{page}} 
\def\bd{
\begin{document}} \def\ed{\end{document}}
\def\bmp{\begin{minipage}} \def\emp{\end{minipage}}
\def\bcc{\begin{center}} \def\ecc{\end{center}}     \def\npg{\newpage}
\def\beq{\begin{equation}} \def\eeq{\end{equation}} \def\hph{\hphantom}
\def\be{\begin{equation}} \def\ee{\end{equation}} \def\r#1{$^{[#1]}$}
\def\n{\noindent} \def\ni{\noindent} \def\pa{\parindent} 
\def\hs{\hskip} \def\vs{\vskip} \def\hf{\hfill} \def\ej{\vfill\eject} 
\def\cl{\centerline} \def\ob{\obeylines}  \def\ls{\leftskip}
\def\underbar#1{$\setbox0=\hbox{#1} \dp0=1.5pt \mathsurround=0pt
   \underline{\box0}$}   \def\ub{\underbar}    \def\ul{\underline} 
\def\f{\left} \def\g{\right} \def\e{{\rm e}} \def\o{\over} \def\d{{\rm d}} 
\def\vf{\varphi} \def\pl{\partial} \def\cov{{\rm cov}} \def\ch{{\rm ch}}
\def\la{\langle} \def\ra{\rangle} \def\EE{e$^+$e$^-$}
\def\bitz{\begin{itemize}} \def\eitz{\end{itemize}}
\def\btbl{\begin{tabular}} \def\etbl{\end{tabular}}
\def\btbb{\begin{tabbing}} \def\etbb{\end{tabbing}}
\def\beqar{\begin{eqnarray}} \def\eeqar{\end{eqnarray}}
\def\\{\hfill\break} \def\dit{\item{-}} \def\i{\item} 
\def\bbb{\begin{thebibliography}{9} \itemsep=-1mm}
\def\ebb{\end{thebibliography}} \def\bb{\bibitem}
\def\bpic{\begin{picture}(260,240)} \def\epic{\end{picture}}
\def\akgt{\noindent{Acknowledgements}}
\def\fgn{\noindent{\bf\large\bf Figure captions}}
\def\EE{e$^+$e$^-$}
\bd
\pge

\vskip-2.5cm
\hskip11cm{\large HZPP-9907}

\hskip11cm{\large July 28, 1999}

\vskip2.5cm

\cl{\Large\bf Identification of Colour Reconnection}
\vs0.2cm
\cl{\Large\bf using Factorial Correlator\footnote{ \ This work is 
supported in part by the
Natural Science Foundation of China \\
\null{} \hs0.6cm (NSFC) under Grant No.19575021.}}

\vs0.8cm
\cl{Fu Jinghua \ \ \ \ \ \ \ \ Liu Lianshou}
\vs0.2cm
\cl{\small  Institute of Particle Physics, Huazhong Normal University,
Wuhan 430079 China}

\cl{\small Tel: 027 87673313 \qquad FAX: 027 87662646
\qquad email: liuls@iopp.ccnu.edu.cn}

\bcc
\begin{minipage}{125mm}
\vskip 1.5cm
\begin{center}{\Large Abstract}\end{center}

\vskip 0.0cm
\ \ \ \ A new signal is proposed for the colour reconnection in the
hadronic decay of W$^+$ W$^-$ in \EE collisions. Using Pythia Monte Carlo
it is shown that this signal, being based on the factorial correlator,
is more sensitive than the ones using only averaged quantities. 

\end{minipage}
\end{center}

\vskip 1in
{\large PACS number: 13.85 Hd
\vskip0.8cm

\ni

\hs2cm Factorial correlator}

\npg \pgn \spg{2}

\vs1.5cm
\def\wp{W$^+$ \hs1pt } \def\wn{W$^-$ \hs1pt }

The discovery of double-W events in the \EE collisions has been a great triumpf
of LEP II experiments. The pure hadronic events, in which the two 
W's decay into four jets, have been ultilized in determining the W mass.

An interesting phenomenon has recently been predicted~\cite{GPZ}\cite{SK}
in connection with these
processes, i.e. the two colour singlets ($q_1 \bar q_2$), ($q_3 \bar q_4$)
formed in the decay of the 2 W's
 $$ {\rm e}^+ {\rm e}^- \rightarrow {\rm W}^+ {\rm W}^- \rightarrow
 q_1\bar q_2 q_3 \bar q_4 \eqno(1) $$
may overlap in space-time, and thus resulting in a colour rearranged state.
This 'colour-reconnection' effect, being a visible dynamics in colour
space, has attracted much attention~\cite{Kittel}. 
It is worthwhile investigation also
because it may affect the accurate determination of W mass~\cite{SK}.

Various signals have been proposed for the identification of this
colour reconnection effect, however, no experimental evidence has ever 
been observed. 

The signals proposed in current literature for this reconnection are the 
shifts in averaged multiplicity $\Delta \la n_{\rm ch}\ra$~\cite{Ellis}, 
in averaged 
scaled-momentum $\Delta \la x_p\ra$~\cite{OPAL}, 
in thrust distribution, etc. No 
significant effect has been observed experimentally using these quantities.
More effective new signal has to be found.

Since colour reconnection is a kind of correlation between two hadronic 
systems formed in the decay of \wp and \wn, it is understandable that the 
averaged quantities as listed above are not sensitive to the effect.
Quantities describing the correlations in multiparticle states are more 
suitable for this purpose. Factorial correlator~\cite{BP}
is a possible candidate.

The factorial correlators are defined as
$$ \la F_{ij}^{mm'}\ra =
\frac
{\la n_m(n_m-1)\cdots(n_m-i+1)n_{m'}(n_{m'}-1)\cdots(n_{m'}-j+1)\ra}
{\la n_m(n_m-1)\cdots(n_m-i+1)\ra\la n_{m'}(n_{m'}-1)\cdots(n_{m'}-j+1)\ra} ,
\eqno(2) $$
where $n_m$ and $n_{m'}$ are the number of particles in the $m$th and
$m'$th bins respectively.
In a fractal system $F_{ij}$ has the anomalous scaling property
$$ \la F_{ij}\ra \propto \f(\Delta/D\g)^{f_{ij}} , \eqno(3) $$
where $D$ is the distance between the two bins.

\def\fc{factorial correlator \ } 
It is well known that the hadronic system (jets) formed in the decay of hard
parton has fractal property~\cite{diffsh}. 
Therefore, we can expect that the \fc of the 
multiparticle
system formed in a single W (either \wp or \wn) will show anomalous scaling
with positive index $f_{ij}$.  
On the other hand, the systems formed from the decay of \wp and \wn are 
independent. If we choose one bin from \wp system and the other bin from 
\wn system, then the correlator $F_{ij}^{+-}$ would be identical to unity
and the corresponding scaling index $f_{ij}^{+-}$ vanishes. 

However, this is only for the case without colour reconnection. When there
were colour reconnection the two systems would not be independent and the 
factorial correlator $F_{ij}^{+-}$ would not be identical to unity.
Let us discuss its possible behaviour in some detail.

In order to get rid of the influence of phase space restriction (boundary 
effect), which will be manifested as a negative correlation 
when the two bins are far away and are both located near by the phase space 
boundaries, we will consider the case of not too large $D$.

As a crude estimation, the colour reconnection can be regarded as an 
exchange of particles between the two systems. Consider for example
the effect when some particles are moved from the \wp system to the \wn
system. This will evidently result in a negative correlation between the
two bins $m^{(+)}$ and $m^{(-)}$ taken from the \wp and \wn systems 
respectively. The nearer is this two bins (the smaller 
is their distance $D$), the stronger will be the negative correlation.
The same will be the case when some particles are moved from the \wn system 
to the \wp system.

Therefore, the factorial correlator $F_{ij}^{+-}$ is expected to be
smaller than unity when there is colour reconnection and the smaller the
distance $D$ is, the smaller is $F_{ij}^{+-}$.

We have used the Pythia event generator to simulate the hadronic decay of WW 
at $\sqrt{s}=184$GeV, which is almost the LEP II energy. Four cases have been
considered. One is the usual case without colour reconnection, the other three
with colour reconnection. The colour reconnection is based on 3 models:

1) SC1, in which the colour field (string) is treated as a Gaussian-profile 
flux tube (similar to a type I superconductor).  The probability of recouple 
is proportional to the overlaping volume. At 184 GeV the recoupling is 
predicted to occur in 38\% of the events.

2) SC2, in which the colour
field is treated as vortex line (similar to type II superconductor). When two
vortex lines cross, they recouple. The recoupling probability at 184 GeV
is 22\% in this case. 

3) SC2p, the same as SC2 but recoupling is allowed only if it reduces the 
string length. The recoupling probability becomes 20\% at 184 GeV.

5000 events have been created for each of the four cases. 

The ``W-axes'' is defined as the direction of 
$\vec p_{{\rm W}^+} - \vec p_{{\rm W}^-}$. The rapidity $y$ is defined as
usual
$$ y= \frac{1}{2} \ln \frac{E+p_\parallel}{E-p_\parallel} , \eqno(4)$$
where $p_\parallel$ is the momentum component along the W-axes.
The analysis is done with final state particles which have energy larger
than 0.1 GeV and non-zero transverse momentum in phase space $-2<y<2$.
The bin-width $\delta y$ is chosen to be 0.1 .

The resulting $F_{11}^{+-}$ are shown in the figure. It can be 
seen that apart from the phase space effect for large $D$ (i.e. ln$D >0$)
the factorial correlators for \wp and \wn without colour reconnection
(full circles) are almost identical to unity, while those for the cases 
with colour reconnection fall down almost linearly in the log-log plot.
This is consistent with the above discussion that the factorial correlattor 
$F_{ij}^{+-}$ is smaller than unity when there is colour reconection and 
the smaller the distance $D$ is, the smaller is $F_{ij}^{+-}$.

Therefore, the behaviour of factorial correlattor $F_{ij}^{+-}$ can
be used as a signal of colour reconnection. If the log $F_{ij}^{+-}$ 
versus -ln$D$ plot in the -ln$D >0$ part falls down from 0, then it 
indicates that there is colour reconnection.

Let us notice that the results of factorial correlator $F_{ij}^{+-}$ 
for the colour-reconnection model SC1 (the 'type I superconductor' model)
behaves a little differently. After an almost linear falling down it turns up
and has a minimum.  
It may possibly be due to the fact that the 'strings' in
this model is extended objects ------ tubes with finite radius ------
instead of thin lines. 
If this  is the case, then this phenomenon may provide a way 
for the differentiation of different reconnection models in the experiments.

In this letter we have shown that the behaviour of factorial correlator
$F_{ij}^{mm'}$ with the two bins $m$ and $m'$ coming from two W's can
be used as a sensitive signal for colour reconnection. This signal, being
based on a correlator, is much more sensitive than the signals proposed
so far, which are based only on averaged quantities.

It should be noticed, however, that the applicability of this method
depends on the assumption that the hadronic systems
from \wp and \wn can be exactly identified. This is hardly to be the case 
using the present jet-algorithm. To develope a better method for the 
partition of final state hadrons into \wp and \wn systems is a challenge 
for further study in 
order that the above-mentioned method to be applicable.
\vskip0.8cm

\akgt \ The authors are grateful to Wes Metzger for encouraging them to do this
investigation.

\npg
\def\J#1#2#3#4{{#1} {\bf #2} (#3) #4}
\def\NCA{\em Nuovo Cimento} \def\NIM{\em Nucl. Instrum. Methods}
\def\NIMA{{\em Nucl. Instrum. Methods} A} \def\NPB{{\em Nucl. Phys.} B}
\def\PLB{{\em Phys. Lett.}  B} \def\PRL{\em Phys. Rev. Lett.}
\def\PRD{{\em Phys. Rev.} D} \def\ZPC{{\em Z. Phys.} C}
\def\PRE{{\em Phys. Rev.} E} \def\PRC{{\em Phys. Rev.} C}

\npg

\begin{picture}(260,240)
\put(-85,-330)
{\epsfig{file=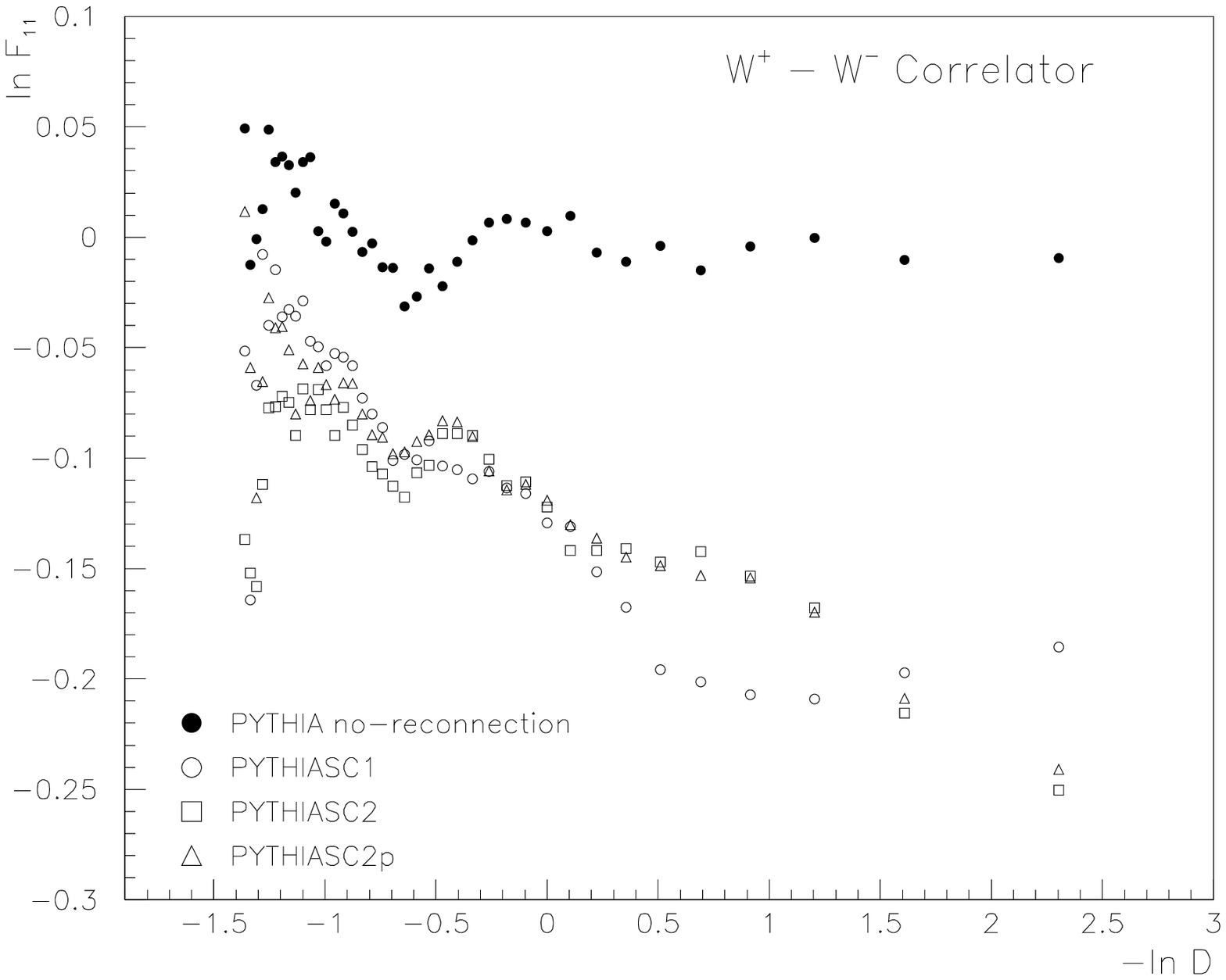,bbllx=0cm,bblly=0cm,
           bburx=8cm,bbury=6cm}}
\end{picture}

\vskip5cm
\cl{
Fig. 1 \ The log-log plot of factorial correlator $F_{11}$ versus $1/D$.
}
\ed